# Spatial Interaction Modelling of Cross-Region R&D Collaborations
## Empirical Evidence from the EU Framework Programmes

Thomas Scherngell[*] and Michael Barber

[*]Corresponding author,
Department of Technology Policy, Austrian Research Centers GmbH – ARC,
Donau-City-Strasse 1, A-1220 Vienna, Austria
Email: thomas.scherngell@arcs.ac.at

August 2008

**Abstract.** The focus of this study is on cross-region R&D collaboration networks in the EU Framework Programmes (FP's). In contrast to most other empirical studies in this field, we shift attention to regions as units of analysis, i.e. we use aggregated data on research collaborations at the regional level. The objective is to identify determinants of cross-region collaboration patterns. In particular, we are interested whether geographical and technological distances are significant determinants of interregional cooperation. Further we investigate differences between intra-industry networks and public research networks (i.e. universities and research organisations). The European coverage is achieved by using data on 255 NUTS-2 regions of the 25 pre-2007 EU member-states, as well as Norway and Switzerland. We adopt a Poisson spatial interaction modelling perspective to analyse these questions. The dependent variable is the intensity of collaborative interactions between two regions, the independent variables are region-specific characteristics and variables that measure the separation between two regions such as geographical or technological distance. The results provide striking evidence that geographical factors are important determinants of cross-region collaboration intensities, but the effect of technological proximity is stronger. R&D collaborations occur most often between organisations that are located close to each other in technological space. Moreover geographical distance effects are significantly higher for intra-industry than for public research collaborations.

# 1  Introduction

Theoretical models of the new growth theory suggest that sustainable economic growth results from increasing returns associated with innovation and the diffusion of new knowledge (see, for example, Romer 1990, Grossman and Helpman 1991). In these models, it is argued that geographical space matters in the innovation process since important parts of new knowledge have some degree of tacitness. Tacit knowledge is embedded in the routines of individuals and organizations. Thus, it is difficult to transfer across geographical space. Theoretical contributions have been followed by a significant body of empirical research on the role of geography in innovation processes, in particular with respect to the geographical dimension of knowledge diffusion and knowledge spillover effects between firms, regions or countries (see, for example, Jaffe, Trajtenberg and Henderson 1993, Zucker et al. 1994, Anselin, Varga and Acs 1997, Almeida and Kogut 1999). In general, by using different statistical and econometric approaches, these studies provide evidence that – as suggested by the theory – knowledge diffusion is geographically localised.

However, the last few years have been characterised by some major changes in the diffusion and production of technological knowledge. The key feature of this shift involves the increasing importance of collaborative network arrangements in the process of knowledge creation and diffusion (see, for example, Castells 1996). The increasing complexity of innovation processes makes it inevitable for innovators to tap external sources of knowledge. In this context, successful innovation depends increasingly on complementary competencies in networks of firms, universities and public research organisations (Ponds, van Oort and Frenken 2007). These changes set up new questions concerning the role of geographical space for innovation and knowledge diffusion since the arrangement of R&D collaboration networks may modify the spatial diffusion of knowledge. Thus, the geographical dimension of innovation and knowledge diffusion deserves further attention by analysing such phenomena as R&D collaborations (Autant-Bernard et al. 2007a).

There have been relatively few empirical studies that investigate the geographical dimension of R&D collaborations[1]. This may be explained by a lack of data on formalised R&D collaboration activities, since networks between individual researchers and between laboratories situated in different settings have long had an informal character. During the recent past such networks have become increasingly formalised and have, thus, received greater institutional visibility[2]. Notable recent contributions involve the studies of Constantelou, Tsakanikas and Caloghirou (2004), Autant-Bernard et al. (2007b), Maggioni, Nosvelli and Uberti (2007) and Maggioni and Uberti (2007). The contribution of Constantelou, Tsakanikas and Caloghirou (2004) investigates inter-country linkages in EU FPs and unveils a picture of significant collaborative activity among clusters of neighboring countries. Autant-Bernard et al. (2007b) compare geographical and social distance effects influencing collaborative patterns in micro- and nanotechnologies, while the study of Maggioni, Nosvelli and Uberti (2007) explain knowledge flows as captured by co-patents as series of exogenous variables including joint participation in FP5 research networks. Maggioni and Uberti (2007) model cross-region collaboration in FP5 programs for five large EU countries with gravity equations estimated by using standard OLS estimation procedures. Generally these studies confirm the relevance of geographical distance for cross-region R&D networks.

The present study is intended to contribute to the existing empirical literature by focusing on cross-region R&D collaboration networks in Europe, as captured by data on research projects of the 5$^{th}$ EU Framework Programme (FP5), and by adopting a Poisson spatial interaction modeling perspective. We take as a working hypothesis that FP5 research projects serve as a proxy for cross-region collaborative R&D activities in

---

[1] It is worth emphasizing in this context that there are empirical studies that investigate the geography of knowledge diffusion, such as the pioneering work by Jaffe, Trajtenberg and Henderson (1993). These studies in general rely on indicators – such as patent citations – that capture knowledge flows between firms, organisations, regions or countries, but they do not explicitly measure formalised R&D collaborations. One recent contribution in this context is the study of Hoekman, Frenken and van Oort (2008) that analyses cross-region research collaboration as measured by scientific publications and patents.

[2] There is an increasing number of empirical studies that investigate structural properties of different European Framework Programme (FP) networks by using social network analysis techniques, such as the contributions of Breschi and Cusmano (2004), Roediger-Schluga and Barber (2006), Protogerou, Caloghirou and Siokas (2007), Billand, Franchisse and Massard (2008) or Roediger-Schluga and Barber (2008). Additionally, Barber et al. (2006) and Almendra et al. (2007) have applied recent methods of statistical physics to networks generated from the FPs.

Europe. We include the 25 pre-2007 EU member-states (Malta and Cyprus excluded), as well as Norway and Switzerland, so that this study extends geographic coverage as compared to most other empirical studies in this field. We shift attention to regions as units of analysis by using aggregated R&D collaboration data at the regional level[3]. Our objective is to identify separation effects on the constitution of cross-region R&D collaborations. In particular, we are interested whether different kinds of geographical space – such as physical distance between regions, existence of country borders between regions or neighbouring region effects – and technological distance between regions are significant determinants of cross-region R&D cooperation in Europe. Further, we are interested in investigating differences between intra-industry networks and public-research networks (i.e. universities and research organisations). Thus, we distinguish between all FP5 collaborations, intra-industry collaborations and intra-public-research collaborations in our modelling approach. A Poisson spatial interaction modelling perspective is applied to address the research questions.

The remainder of this study is organised as follows. *Section 2* provides some further insight on R&D collaborations and the EU FPs that have been established to foster collaborative R&D activities in Europe. It sheds some light on the thematic priorities and the formal application rules of FP5. *Section 3* presents the empirical model used to investigate cross-region R&D collaborations by shifting attention to the spatial interaction modelling perspective. *Section 4* discusses in some detail the data used and describes the construction of the dependent and the independent variables accompanied by some descriptive statistics on the nature of cross-region R&D collaborations in Europe. *Section 5* continues to describe model specification, adopting a Poisson specification where model parameters are derived from Maximum Likelihood estimation. *Section 6* presents the estimation results, while *Section 7* concludes with a summary of the main results.

---

[3] As pointed out by the European Commission (2001) and Lagendijk (2001), regions are central sites for knowledge creation as well as for knowledge diffusion and learning processes in the new age of the knowledge-based economy and are becoming increasingly important as policy units for research and innovation.

## 2 R&D collaborations and the European Framework Programmes

It is widely believed that interaction between firms, universities and research organisations is a sine-qua-non condition for successful innovation in the current era of the knowledge-based economy, in particular in knowledge intensive industries, and, thus, for sustained economic competitiveness (see, for example, OECD 1992). Pavitt (2005) notes that the growing complexity of technology and the existence of converging technologies are key reasons for this development. In particular, firms have expanded their knowledge bases into a wider range of technologies (Granstrand 1998), which increases the need for more different types of knowledge, so firms must learn how to integrate new knowledge into existing products or production processes (Cowan 2004). It may be difficult to develop this knowledge alone or acquire it via the market. Thus, firms aim to form different kinds of cooperative arrangements with other firms, universities or research organisations that already have this knowledge to get faster access to it. The study of Hagedoorn and Kranenburg (2003) confirms the rise of strategic R&D alliances during the 1990s.

The fundamental importance of interactions and networks for innovations is also reflected in the various systems of innovation concepts (see Lundvall 1992, among others). In this conception, the sources of innovation are often established between firms, universities, suppliers and customers. Network arrangements create incentives for interactive organisational learning leading to faster knowledge diffusion within the innovation system and stimulating the creation of new knowledge or new combinations of existing knowledge. In particular, network arrangements are useful in the presence of uncertainty and complexity, such as in innovation processes. Participation in innovation networks reduces the degree of uncertainty and provides fast access to different kinds of knowledge, in particular tacit knowledge (see, for example, Kogut 1988)[4].

---

[4] Other theoretical contributions concerning motives for network arrangments include transaction cost economics (Williamson 1975) or the resource-based view (Penrose 1959).

Over the last few years the EU has followed the systems of innovation view with respect to the strategic orientation of its technology and innovation policies[5]. The main instrument in this context are the Framework Programmes (FPs) on Research and Technological Development, which have funded thousands of collaborative R&D projects to support transnational cooperation and mobility for training purposes. Based on the Maastricht treaty of the EU, the FPs were implemented to realise two main objectives: *First*, strengthening the scientific and technological bases of industry to foster international competitiveness and, *second*, the promotion of research activities in support of other EU policies (CORDIS 2006). Implementation of the EU FPs began in 1984; the current seventh programme has begun in 2007 and will run until 2013[6]. There is evidence from some exploratory studies that the EU FPs have a major impact on the formation of networks in Europe (see Breschi and Cusmano 2004). Roediger-Schluga and Barber (2006) show that the integration between collaborating organisations has increased over time by using social network analysis techniques and conclude that these findings point to a trend towards a more integrated European Research Area.

Since the launch of the FPs in 1984, EU institutions have focused funding of multidisciplinary research at a trans-national level. Over the years, different thematic aspects and issues of the European scientific landscape have been addressed by the FPs though the main emphasis shifted more and more towards the establishment of an integrated European Research Area (see Kruckenberg, Brandes and Ahrweiler 2008). In this sense, the European technology policy aims to promote technological competitiveness while at the same time it is meant to ensure cohesion (see Peterson and Sharp 1998). Projects to be funded by the FPs need to fulfil a number of governance rules that shape the particular outset of each FP and present a formal framework of participation. One of the key governance objectives over all FPs is to support scientific work of the highest quality. In this context, the European scientific community should benefit from mobile researchers being involved in transnational projects (see Kruckenberg, Brandes and Ahrweiler 2008).

---

[5] See Caloghirou, Vonortas and Ioannides (2002) for a detailed discussion on other major international examples.

[6] See Roediger-Schluga and Barber (2006) for a detailed discussion on the history and different scopes of the EU FPs since 1984.

In spite of their different scopes, the fundamental rationale of the FPs has remained unchanged (see Barker and Cameron 2004). The current study uses data from FP5 projects in order to capture cross-region collaborative activities. In 1998, the European Council authorised a budget of 13.7 billion EUR for FP5 (CORDIS 1998). Within FP5 research projects were funded within a time interval of five years (1998-2002). In reaction to a rising number of critiques on FP4 – in particular with respect to the negligence of marketability, support of Small and Medium Enterprise (SME) and slow and bureaucratic procedures for payments and proposal selection (see Peterson and Sharp 1998) – the overall structure of FP5 turned out to be slightly modified (see Kruckenberg, Brandes and Ahrweiler 2008). FP5 focused on a limited number of research areas combining technological, industrial, economic, social and cultural aspects[7]. Furthermore, FP5 shifted emphasis on the protection of intellectual property rights. By this, the European Commission tried to improve the efficiency of collaboration within the various types of European research projects. In addition, new project management procedures were introduced. For instance, the role of project coordinators became increasingly important because new administrative duties with respect to scientific coordination, supervision, and reports on financial issues were now assigned to them (see Kruckenberg, Brandes and Ahrweiler 2008).

Project proposals in FP5 are submitted by self-organised consortia (see European Council 1998). There are several formal criteria that should be noted in this study. FP5, with its corresponding financial support, is open to all legal entities established in the Member States of the European Union – e.g. individuals, industrial and commercial firms, universities, research organisations, etc. (see CORDIS 1998). Proposals can be submitted by at least two independent legal entities established in different EU Member States or in an EU Member State and an Associated State[8]. Proposals to be funded are selected on the basis of criteria including scientific excellence, added value for the European Community, the potential contribution to furthering the economic and social objectives of the Community, the innovative nature, the prospects for disseminating/exploiting the results, and effective transnational cooperation. Further details on the formal application procedures are given by the European Council (1998).

---

[7] The thematic priorites in FP5 are the following (Subprogramme name given in brackets): Quality of Life and management of living resources (Quality of Life); User-friendly information society (IST); Competitive and sustainable growth (GROWTH); Energy, environment and sustainable developement (EESD); Confirming the international role of community resaerch (INCO2); Promotion of innovation and encouragement of SME participation (Innovation/SMEs); Improving the human reserach potential and the socio-economic knowledge base (Improving) (CORDIS 1998).

[8] Associated States include the canditates of EU-membership in this time period (Bulgaria, Cyprus, Czech Republic, Estonia, Hungary, Latvia, Lithuania, Malta, Poland, Romania, Slovakia, Slovenia) as well as Iceland, Isreal, Liechtenstein, Norway and Switzerland. For more details see CORDIS (1998).

## 3  The empirical model of cross-region R&D collaborations in Europe

In our analytical framework we adopt a spatial interaction modelling perspective. In this section, we present the modelling framework at an abstract level. Construction of the model variables and their concrete interpretations are presented in Section 4.

We model cross-region R&D collaborations to examine how specific separation effects explain the variation of cross-region R&D collaborations in Europe. Denoting regions by $i, j = 1,\ldots, n$ and letting $\boldsymbol{P}$ be the $n$-by-$n$ square matrix of observed cross-region R&D collaborations, where the element $p_{ij}$ is the observed number of R&D collaborations between two regions $i$ and $j$, the basic model takes the form

$$P_{ij} = V_{ij} + \varepsilon_{ij} \qquad\qquad i, j = 1,\ldots, n \qquad (1)$$

where $P_{ij}$ is a stochastic dependent variable that corresponds to observed R&D collaborations $p_{ij}$, with the property $E[P_{ij} \mid p_{ij}] = V_{ij}$. $V_{ij}$ denotes the systemic part of the model that captures the stochastic relationship to other model variables, which are the covariates. $\varepsilon_{ij}$ is a random term that varies across all ($i$, $j$)-region pairs, with $E[\varepsilon_{ij} \mid p_{ij}] = 0$ as a minimal requirement.

An appropriate model for $V_{ij}$ is the spatial interaction model of the gravity type. Spatial interactions models can be grouped under the generic heading gravity models (Roy and Thill 2004). They have gained wide acceptance as a reasonable model of spatial interactions between locations (such as regions). Spatial interaction models incorporate a function characterising the origin *i* of interaction, a function characterising the destination *j* of interaction and a function characterising the separation between two regions *i* and *j*. The model is characterised by a formal distinction implicit in the definitions of origins and destination functions on the one hand, and separation functions on the other (see, for example, LeSage, Fischer and Scherngell 2007). Origin and destination functions are described using weighted origin and destination variables, respectively, while the separation functions are postulated to be explicit functions of numerical separation variables[9]. Thus, the spatial interaction model is given by

$$V_{ij} = A_i \ B_j \ S_{ij} \qquad\qquad i,j = 1,\ldots,n \qquad (2)$$

for some appropriate choice of an origin function, $A_i$, destination function, $B_j$, and separation function, $S_{ij}$. In this study we follow classical spatial interaction model specifications and define $A_i = A(a_i, \alpha_1) = a_i^{\alpha_1}$, and $B_j = B(b_j, \alpha_2) = b_j^{\alpha_2}$ where $a_i$ and $b_j$ are mass terms denoting some appropriate origin and destination variables, respectively. $\alpha_1$ and $\alpha_2$ are scalar parameters to be estimated. The product of the functions $A_i B_j$ in Equation (2) can be simply interpreted as the number of cross-region R&D collaborations which are possible.

With respect to the research questions of the current study, the focus of interest is on the separation function $S_{ij}$ that constitutes the very core of spatial interaction models. $S_{ij}$ is hypothesised to summarise all effects of geographical and technological space on cross-region R&D collaborations. For our emphasis on the geography of R&D collaborations in Europe, the specification of $S_{ij}$ is of central importance. Again we have chosen to follow spatial interaction theory and use a multivariate exponential functional form that is given by

$$S_{ij} = S(d_{ij}, \beta) = \exp\left[\sum_{k=1}^{K} \beta_k \ d_{ij}^{(k)}\right] \qquad\qquad i,j = 1,\ldots,n \qquad (3)$$

---

[9] See Sen and Smith (1995) for a theoretical underpinning in spatial interaction theory and analysis.

where $d_{ij}^{(k)}$ are $K$ separation measures and $\beta_k$ ($k = 1, ..., K$) are parameters to be estimated. For the purposes of this study this class of multivariate separation functions provides a very flexible representational framework.

We focus on $K = 6$ distinct measures of separation. $d_{ij}^{(1)}$ denotes geographical distance between two regions $i$ and $j$. $d_{ij}^{(2)}$ captures country border effects to test if a country border between two regions $i$ and $j$ affects cross-region R&D collaborations, while $d_{ij}^{(3)}$ accounts for language barrier effects on collaborative activities. $d_{ij}^{(4)}$ measures the distance in technological space between two regions $i$ and $j$ in order to capture the impact of technological effects on collaborative activities in the EU FPs. $d_{ij}^{(5)}$ and $d_{ij}^{(6)}$ control for neighbouring region- and neighbouring country effects. They are included to estimate how co-localisation of organisations in neighbouring regions/countries affects the likelihood that they collaborate. The specification of these variables is discussed in some detail in *Section 4*.

Integrating the origin, destination and separation functions into Equation (1) leads to the empirical model:

$$p_{ij} = a_i^{\alpha_1} b_j^{\alpha_2} \exp\left[\sum_{k=1}^{K} \beta_k \, d_{ij}^{(k)}\right] + \varepsilon_{ij}. \tag{4}$$

We are interested in estimating the parameters $\alpha_1$, $\alpha_2$ and $\beta_k$ that are elasticities of cross-region R&D collaborations $p_{ij}$ with respect to the origin variable $a_i$, the destination variable $b_j$ and the separation variables $d_{ij}^{(k)}$.

## 4  Data, variable definition and some descriptive statistics

This section discusses in some detail the empirical setting and the construction of the dependent and the independent variables. For the construction of different cross-region

R&D collaboration matrices $P$ and the origin and destination variables $a_i$ and $b_j$ we use data on R&D collaborations funded within FP5, while we draw on geographical information systems (GIS) data and patent applications assigned at the European patent office (EPO) for the separation variables $d_{ij}^{(k)}$. The European coverage is achieved by using cross-section data on $i, j = 1, …, n = 255$ NUTS-2 regions[10] (NUTS revision 2003) of the 25 pre-2007 EU member-states, as well as Norway and Switzerland. The detailed list of regions is given in Appendix A.

**The region-by-region R&D collaboration matrices**

Our core data set to capture collaborative activities in Europe is the *sysres EUPRO* database[11] that presently comprises data on funded research projects of the EU FPs (complete for FP1-FP5, and about 70% for FP6) and all participating organisations. It contains systematic information on project objectives and achievements, project costs, project funding and contract type as well as on the participating organisations including the full name, the full address and the type of the organisation. We use a concordance scheme between postal codes and NUTS regions provided by Eurostat to trace the specific NUTS-2 region of an organisation. Thus, the *sysres EUPRO* database represents an extremely valuable resource not only for this study, but for any kind of empirical analysis on the geography of knowledge creation and diffusion across Europe.

To construct our region-by-region collaboration matrices $P$ we aggregate the number of individual collaborative activities to the regional level which leads to the observed number of R&D collaborations $p_{ij}$ between two regions $i$ and $j$. We make use of the respective NUTS-2 regions of 23,318 organisations participating in 9,456 projects of FP5[12]. For instance, for a project with three participating organisations in three different regions, say regions *a*, *b*, and *c*, we count three links: from region *a* to region *b*, from *b* to *c* and from *a* to *c*[13]. When all three participants are located in one region we count three intraregional links. Note that we have excluded self loops to eliminate artificial self collaborations. The resulting regional collaboration matrix ***P*** then contains the total collaboration intensities between all (*i*, *j*)-region pairs, given the $i = 1, \ldots, n = 255$ regions in the rows and the $j = 1, \ldots, n = 255$ regions in the columns. The *n*-by-*n* matrix is symmetric by construction ($p_{ij} = p_{ji}$).

---

[10] NUTS is an acronym of the French for the "nomenclature of territorial units for statistics", which is a hierarchical system of regions used by the statistical office of the European Community for the production of regional statistics. At the top of the hierarchy are NUTS-0 regions (countries) below which are NUTS-1 regions and then NUTS-2 regions. Although varying considerably in size, NUTS-2 regions are widely viewed as the most appropriate unit for modelling and analysis purposes (see, for example, Fingleton 2001).

[11] The *sysres EUPRO* database is constructed and maintained by ARC systems research by substantially standardising raw data on EU FP research collaborations obtained from the CORDIS database (see Roediger-Schluga and Barber 2008).

Furthermore, ***P*** is the basis for constructing two collaboration matrices ***P*$^{(ind)}$** and ***P*$^{(edu)}$** that we need to identify differences between intra-industry and intra-public-research collaboration activities. ***P*$^{(ind)}$** is extracted from ***P*** by excluding all non-industry organisations, while ***P*$^{(edu)}$** is extracted from ***P*** by excluding all non-public-research organisations from our dataset. Note that the dimensions of the region-by-region collaboration matrices ***P*$^{(ind)}$** and ***P*$^{(edu)}$** are the same as of ***P***.

As a prelude to the analysis that follows in the next sections, Table 1 presents some descriptive statistics on total R&D collaborations among the 255 (*i*, *j*)-region pairs as given by ***P***. There are about 730,000 FP5 cross-region collaborations. The mean number of collaborations between any two regions is 11.12, with standard deviation 46.83. About 40% of all pairs of regions (25,211 pairs) do not collaborate at all. The mean collaboration intensity for region pairs that have at least one joint research project is 18.32. 33,288 links are intraregional ones and found on the main diagonal of the matrix. The mean intraregional collaboration intensity is 130.66 and much higher than the mean interregional collaboration intensity (10.72). The off-diagonal elements show an extremely right-skewed distribution (the median is 1, the mode is zero).

---

[12] We use FP5 since data on FP6 are not complete at the current stage of *sysres EUPRO*. FP5 ran from 1998-2002. We include all types of collaborative activities, i.e. all projects that have at least two members. We assume our data to represent a "paper trail" recording evidence of scientific and technological cooperation and networking between regions in the period of 1998-2002. We note that we are uninterested in analysing rejected consortia proposals since these may not lead to cross-region R&D activities.

[13] We refer to this counting method as full counting. Another counting method would be fractional counting by dividing each link in a project by the total number of links in a project. The full counting procedure used in the current study overestimates the impact of large projects, while the impact of large projects would be underestimated using the fractional counting method. We prefer the full counting method since full rather than fractional counting does justice to the true integer nature of R&D collaborations and is applicable in the context of a Poisson model specification.

**Table 1: Some descriptive statistics on R&D collaborations among European regions as captured by joint EU FP5 research projects**

|  | Matrix Elements | Sum | Mean | Standard Deviation | Min | Max |
|---|---|---|---|---|---|---|
| **All Links** | 65,025 | 728,120 | 11.12 | 46.83 | 0 | 6,152* |
| **Positive Links** | 39,743 | 728,120 | 18.32 | 39.12 | 1 | 6,152* |
| **Intraregional Links** | 255 | 33,288 | 130.66 | 443.35 | 0 | 6,152* |
| **Interregional Links** | 64,700 | 694,832 | 10.72 | 37.07 | 0 | 1,609** |
| **Positive Interregional Links** | 39,489 | 694,832 | 17.57 | 46.12 | 1 | 1,609** |
| **National Interregional Links** | 4,976 | 75,536 | 15.12 | 43.81 | 0 | 1,072*** |
| **International Interregional Links** | 59,794 | 619,296 | 10.34 | 36.43 | 0 | 1,609** |

*within Île-de-France, **between Île-de-France and Oberbayern, ***between Île-de-France and Rhone-Alpes

Table 1 also indicates that intranational R&D collaborations are more frequent than international ones; the mean collaboration intensity for intranational collaborations is 15.12, while for international it is 10.34. The difference between intra- and international collaboration frequency may point to the existence of country border effects, which will be tested in the spatial interaction model. The maximum collaboration intensity is 6,152 referring to intraregional collaborations within the region of Île-de-France, while the maximum interregional collaboration activity (1,609) includes R&D collaborations between Île-de-France and Oberbayern (Germany). The largest interregional collaboration activity within one country is found for the region pair Île-de-France and Rhone-Alpes (1,072 collaborations).

Figure 1 illustrates the skewness of R&D collaborations across European regions using a histogram. The frequency of collaborative activities declines very quickly for more intensive collaboration links. Relatively few region pairs show a high number of R&D collaborations, while the majority of the region pairs (more then 45,000) show a collaboration intensity lower than 11. There are only 1,112 region pairs for which the number of collaborations is over 100.

**Figure 1: Frequency of total cross-region R&D collaborations in Europe**

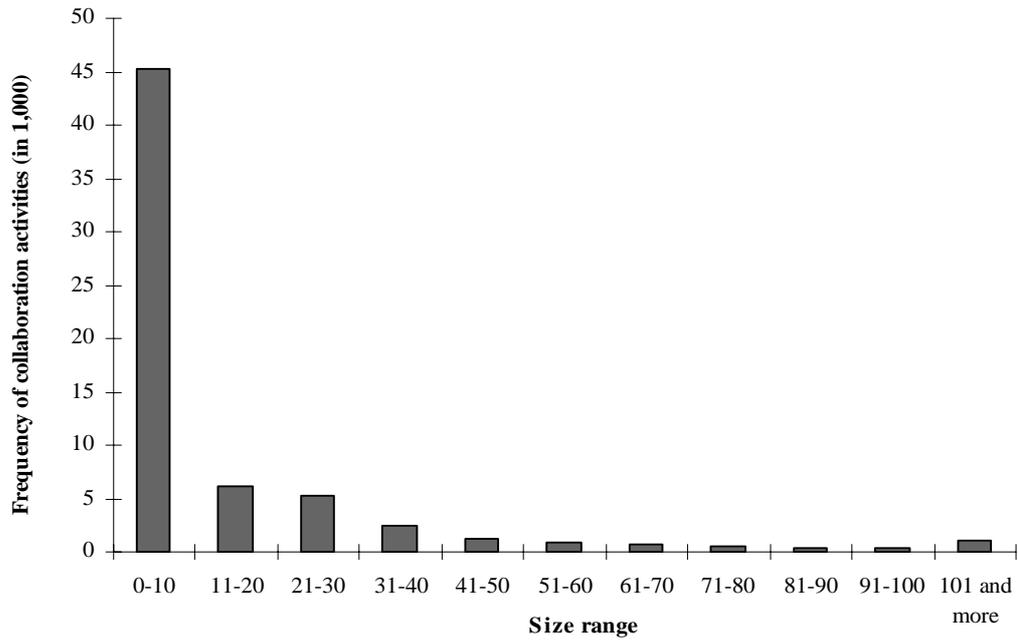

**Figure 2: Total cross-region R&D collaborations in Europe as captured by research projects funded by EU FP5**

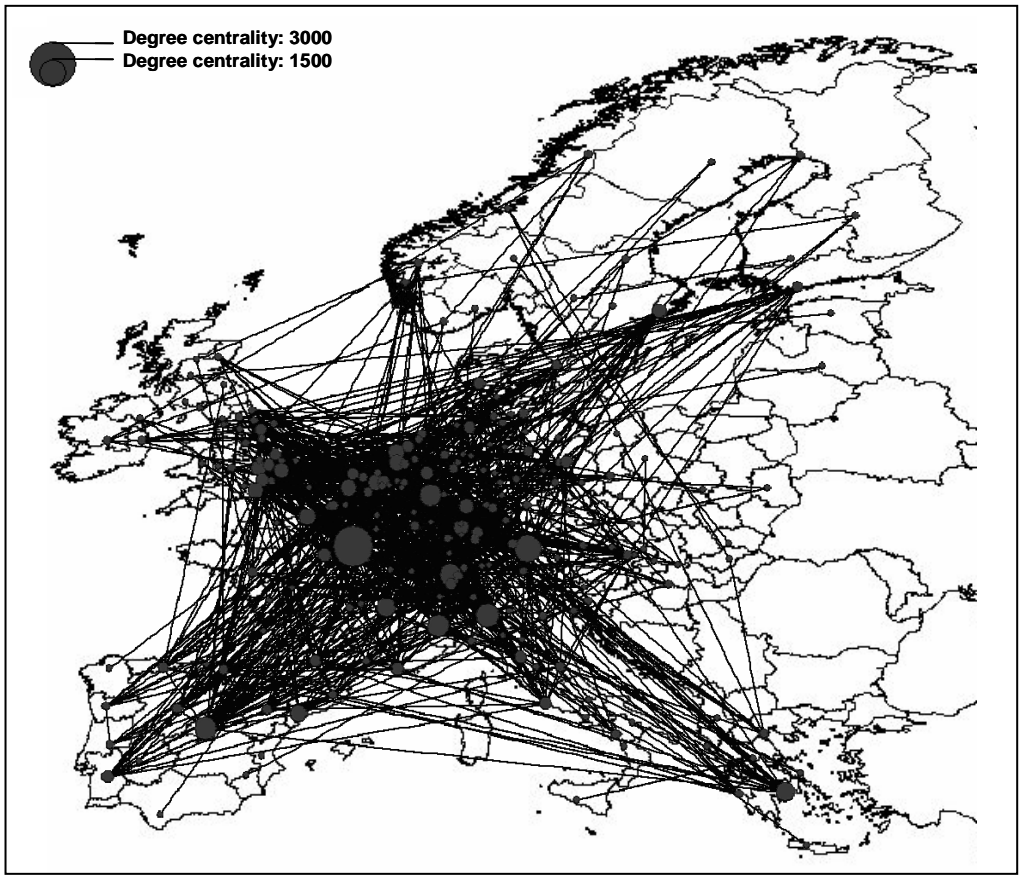

The spatial region-by-region R&D network is visualised in Figure 2. The nodes represent one region; their size is relative to their degree centrality corresponding to the number of links connected to a region[14]. The central hub in this spatial network is Île-de-France, a high density can also be observed for southeastern regions of the UK, northern Italian regions, southern and western regions in Germany, the Netherlands and Switzerland as well as for the capital regions in Greece and Spain. The number of links to eastern European regions is generally quite low.

**The independent variables**

We again draw on data from the *sysres EUPRO* database for the origin variable $a_i$ and the destination variable $b_j$, respectively. The origin variable is simply measured in terms of the number of organisations participating in EU FP5 projects in the region *i*, while the destination variable denotes the number of organisations participating in EU FP5 projects in region *j*. Note that the values for $a_i$ and $b_j$ are the same, but their interpretation in the spatial interaction modelling estimation is different.

All separation variables with the exception of $d_{ij}^{(4)}$ are identified from geographical information systems data. We use the great circle distance between the economic centres of two regions *i* and *j* to measure the geographical distance variable $d_{ij}^{(1)}$. $d_{ij}^{(2)}$ is a country border dummy variable that takes a value of zero if two regions *i* and *j* are located in the same country, and one otherwise, while $d_{ij}^{(3)}$ is a language area dummy variable that takes a value of zero if two regions *i* and *j* are located in the same language area, and one otherwise[15]. $d_{ij}^{(5)}$ and $d_{ij}^{(6)}$ are dummy variables that take a value of one if the regions *i* and *j* are direct neighbours[16] or are located in neighbouring countries, respectively, and zero otherwise.

---

[14] Note that the region-by-region network is an undirected graph from a network analysis perspective.

[15] Language areas are defined by the region`s dominant language (see LeSage, Fischer and Scherngell 2007). However, in most cases the language areas are combined countries, as for instance Austria, Germany and Switzerland (one exception is Belgium, where the French speaking regions are separated from the Flemish speaking regions).

[16] We define two regions *i* and *j* as neighbors when they share a common border.

To define the technological distance $d_{ij}^{(4)}$ between two regions $i$ and $j$ we use data on European patent applications from the European patent office (EPO) database that have an application date between 1998 and 2003[17]. The variable is constructed as a vector $t(i)$ that measures region $i$'s share of patenting in each of the technological subclasses of the International Patent Classification (IPC). Technological subclasses correspond to the third-digit level of the IPC systems. We use the Pearson correlation coefficient given by $r^2 = corr[t(i), t(j)]^2$ between the technological vectors of two regions $i$ and $j$ to define how close they are to each other in technological space. Their technological distance is given by $d_{ij}^{(4)} = 1 - r^2$ (see Moreno, Paci and Usai 2005).

## 5  The Poisson model specification with heterogeneity

At this point we would like to estimate the parameters of our empirical model of cross-region R&D collaborations given by Equation (4). At a first glance it is tempting to express model (4) equivalently as a log-additive model and estimate the parameters using ordinary or non-linear least squares procedures (see, for example, Bergkvist and Westin 1997). However, this approach suffers from three drawbacks since we model count data: *First*, taking logarithms on both sides of Equation (4) would model discrete count outcomes by a continuous process that is misrepresentative. *Second*, a logarithmic form of model (3) estimated by OLS or NLS assumes $\varepsilon \sim N(0, \sigma^2)$ which would only be justified statistically if the collaborations $p_{ij}$ were log-normally distributed with a constant variance. This is not a natural assumption in our case due to the discrete count data process of $p_{ij}$ (see also Figure 1). *Third*, the logarithm of zero is not defined but the fraction of zeros in the dependent variable is about 40%[18].

---

[17]  Patents have been used widely in the scientific literature to capture knowledge outputs. They provide a very rich and useful source of data for the study of innovation and technological change. See, for example, Griliches (1990) for a detailed discussion on the advantages and shortcomings of patent data.

[18]  Some studies exclude the zero flows during estimation or replace them by a very small flow. In the first case information is decreased while in the second case disinformation is increased neither of which are very satisfying (see Bergkvist and Westin 1997). Aggregation of the network until all flows are positive is also not possible in our case because then we would have to aggregate the collaboration flows to the level of countries which is inappropriate for the focus of the current study.

To overcome the problems of least squares assumptions, it is suitable to use a Poisson model specification. The Poisson distribution is generally considered as a reasonable description for non-negative integer values, in particular in the case of rare events (see, for example, Kennedy 2003). The Poisson distribution provides the probability of the number of event occurrences in the model, and the Poisson parameters corresponding to the expected number of occurrences are modelled as a function of explanatory variables. Also the Poisson specification of the model has no problems with the zero flows since $p_{ij} = 0$ is a natural outcome of the Poisson process. The Poisson density function is given by

$$f(p_{ij}) = V_{ij}^{p_{ij}} e^{-V_{ij}} / p_{ij}! \tag{5}$$

with

$$V_{ij} = A(a_i, \alpha_1) B(b_j, \alpha_2) S(d_{ij}^{(k)}, \beta_k) \tag{6}$$

The standard estimator for the Poisson spatial interaction model is the maximum likelihood estimator (see Fischer, Scherngell and Jansenberger 2006 for details on the ML estimation).

The Poisson model specification assumes

$$V_{ij} = Var\left[p_{ij} | A_i, B_j, S_{ij}\right] = E\left[p_{ij} | A_i, B_j, S_{ij}\right], \tag{7}$$

which implies that the independent variables account for all individual deviations. Thus, model (5) with (6) may suffer from unobserved heterogeneity between the $(i, j)$-region pairs. Since we deal with a multiregional setting the existence of unobserved heterogeneity is very likely. Unobserved heterogeneity that cannot be captured by the covariates may lead to biased estimates due to overdispersion, i.e. assumption (7) does not hold true. As noted by, for instance, Long and Freese (2001), a very promising strategy to overcome the problem of unobserved heterogeneity is to introduce a stochastic heterogeneity parameter $\exp(\xi_{ij})$ leading to a modification of Equation (6) by

$$V_{ij}^* = \exp\left[\log A(a_i, \alpha_1) + \log B(b_j, \alpha_2) + \log S(d_{ij}, \beta_k) + \xi_{ij}(\delta)\right] \qquad (8)$$

where $\delta$ is the dispersion parameter. When $\xi_{ij} \sim Gamma$, then $p_{ij} \sim Negative$ *Binomial*, leading to a Negative Binomial density distribution that is given by (see Long and Freese 2001)

$$f(p_{ij}) = \frac{\Gamma(p_{ij} + \delta^{-1})}{\Gamma(p_{ij} + 1)\Gamma(\delta^{-1})} \left(\frac{\delta^{-1}}{V_{ij} + \delta^{-1}}\right)^{\delta^{-1}} \left(\frac{V_{ij}}{V_{ij} + \delta^{-1}}\right)^{p_{ij}} \qquad (9)$$

where $\Gamma(\cdot)$ denotes the gamma function. The model allows for overdispersion $\delta > 0$ by $Var(V_{ij}) = E(V_{ij}^*) + Var(V_{ij}^*)$ [19]. Note that when $\delta = 0$, model (9) collapses to the standard Poisson specification without heterogeneity. Model estimation is again done by Maximum Likelihood (see Cameron and Trivedi 1998).

## 6 Estimation results

Table 2 presents the sample estimates of the spatial interaction models, with standard errors given in brackets. We use the Negative Binomial model specification as given by Equations (8) and (9). The dispersion parameter $\delta$ is significant for all model versions, indicating that the Negative Binomial specification is to be preferred over the standard Poisson model without heterogeneity. The existence of unobserved heterogeneity that cannot be captured by the covariates leads to overdispersion and, thus, to biased model parameters for the Poisson model without heterogeneity. We estimate three model version: Model (1) uses the total cross-region FP5 R&D collaborations as given by *P* as the dependent variable, model (2) the intra-industry cross-region FP5 R&D collaborations as given by $P^{(ind)}$, and model (3) the intra-public-research cross-region FP5 R&D collaborations as given by $P^{(edu)}$.

---

[19] We rejected using zero-inflated Poisson/Negative Binomal specifications since none of the region pairs is excluded from collaboration in principle, even though the fraction of zeros is about 40%. However, we calculated zero-inflated Negative Binomial specification as given in Long and Freese (2001) to compare statistical fit. The respective Vuong Test yields a value of -1.67 which is not significant, i.e. the statistical fit of the zero-inflated model is not significantly higher (see Long and Freese 2001). Thus, we prefer the standard Negative Binomial model due to a better theoretical underpinning.

**Table 2: Estimation Results of the Negative Binomial Spatial Interaction Models**
[65,025 observations, asymptotic standard errors given in brackets]

|  | Negative Binomial spatial interaction model | | |
| --- | --- | --- | --- |
|  | (1) Total FP5 | (2) FP5 Industry | (3) FP5 University |
| *Origin variable* [$\alpha_1$] | 0.973*** (0.002) | 0.922*** (0.006) | 0.996*** (0.003) |
| *Destination variable* [$\alpha_2$] | 0.974*** (0.002) | 0.921*** (0.006) | 0.997*** (0.003) |
| *Geographical distance* [$\beta_1$] | -0.228*** (0.005) | -0.340*** (0.013) | -0.077*** (0.007) |
| *Country border effects* [$\beta_2$] | -0.048** (0.017) | -0.038* (0.014) | -0.002 (0.004) |
| *Language area effects* [$\beta_3$] | -0.119*** (0.015) | -0.082* (0.037) | -0.057** (0.019) |
| *Technological distance* [$\beta_4$] | -0.677*** (0.071) | -0.823*** (0.104) | -0.721*** (0.097) |
| *Neighbouring region* [$\beta_5$] | 0.256*** (0.022) | 0.325*** (0.052) | 0.111*** (0.029) |
| *Neighbouring country* [$\beta_6$] | 0.080*** (0.009) | 0.040** (0.001) | 0.035* (0.011) |
| **Constant** | -6.131*** (0.077) | -7.491*** (0.195) | -8.319*** (0.096) |
| **Dispersion parameter** ($\delta$) | 4.271*** (0.051) | 0.729*** (0.105) | 3.610*** (0.050) |
| **Log-Likelihood** | -126,729.12 | -123,561.04 | -124,823.70 |
| **Sigma Square** | 8.823 | 7.591 | 7.976 |

Notes: The dependent variable in model (1) is the cross-region collaboration intensity between two regions *i* and *j*, in model (2) the cross-region intra-industry collaboration intensity between two regions *i* and *j*, and in model (3) the cross-region intra-public-research collaboration intensity between two regions *i* and *j*. The independent variables are defined as given in the text. Note that we tested the residual vector for the existence of spatial autocorrelation which could be a problem in the context of interaction data (see LeSage, Fischer and Scherngell 2007). The respective Moran´s *I* statistic is insignificant, i.e. spatial autocorrelation in the error term does not exist. ***significant at the 0.001 significance level, **significant at the 0.01 significance level, *significant at the 0.05 significance level

In general the parameter estimates are robust and highly significant over all model versions. In the context of the relevant literature on innovation and knowledge diffusion our model produces some interesting results. Model (1) provides evidence that geographical distance between two organisations has a significant negative effect on the likelihood that they collaborate. The parameter estimate of $\beta_1 = -0.228$ indicates that for each additional 100 km between two organisations, the mean collaboration frequency decreases by 25.6%. However, the effect of geographical distance becomes even more important when we look at intra-industry collaborations as evidenced by model (2). $\beta_1$ increases in magnitude by 67% to a value of -0.340. In contrast, model (3) shows that geographical distance has a much lower negative effect for intra-public-research collaborations. The effect of geographical distance nearly disappeared in the case of intra-public-research collaboration activities.

Country borders have – as evidenced by the estimate for $\beta_2$ – a comparatively small effect. This is unsurprising since EU FP projects must have at least one international partner (*see Section 2*). However, there is still a small negative effect observable that is significantly different from zero for total FP5 collaborations (model (1)) and intra-industry collaborations (model (2)). As given by model (3), country border do not affect the probability for public research organisations to cooperate with each other.

The estimate for $\beta_3$ tells us that it is more likely that collaborations occur between regions that are located in the same language area, but the effect of language barriers is smaller than geographical distance effects. For all model versions, most important are technological distance effects as evidenced by the parameter estimates for $\beta_4$. This implies that it is most likely that cross-region R&D collaborations occur between regions that are close to each other in technological space. This finding is in line with previous results of Fischer, Scherngell and Jansenberger (2006) and LeSage, Fischer and Schengell (2007) for the case of interregional knowledge spillovers, but the technological distance effect they found is much higher than in the current study for interregional FP collaborations.

The parameter estimates of $\beta_5$ and $\beta_6$ indicate that the likelihood of collaboration between two organisations increases when they are located in neighbouring regions or neighbouring countries, respectively. Neighbouring region effects are somewhat larger than geographical distance effects but also smaller than technological distance effects. As for geographical distance, neighbouring region/country effects are highest for intra-industry collaborations, while for intra-public-research collaborations they are quite small. As expected, the estimates for the origin and destination variables $\alpha_1$ and $\alpha_2$ are close to one for all model versions, indicating that a higher number of participating organisations in a region increases the probability of collaboration with other regions. Further, $\alpha_1$ and $\alpha_2$ are equal up to numerical precision, as we expect from the assumed symmetric interactions.

# 7  Concluding remarks

One of the key current research fields in economic geography and economics of innovation is the empirical analysis of the geography of innovation. In particular, the geographical dimension of phenomena such as R&D collaborations is of special interest in order to gain insight into the spatial diffusion of knowledge. The analysis of the geography of R&D collaboration has important policy implications for the EU, for instance with respect to the spatial scale of innovation systems and R&D interactions.

The focus of this study is on the role of geographical space for R&D collaborations funded by the 5$^{th}$ EU FP from a regional perspective. The objective was to identify separation effects – including different kinds of geographical space and technological distance – on the constitution of cross-region R&D collaborations. We used an appropriate analytical framework, the Poisson spatial interaction modelling perspective, to estimate the separation effects.

The study produces some promising results in the context of the empirical literature on innovation. Geographical distance and co-localisation of organisations in neighbouring regions are important determinants of the constitution of cross-region R&D collaborations in Europe. However, model estimations show that these geographical effects are much higher for intra-industry cooperative activities than for collaborations between public research organisations, where negative effects of geography nearly vanish.

For all model versions, technological proximity is more important than spatial effects. R&D collaborations occur most often between organisations that are not too far from each other in technological space. R&D collaborations are also determined by language barriers, but language barrier effects are smaller than geographical effects. Country border effects are rather small, due to the governance rule implemented by the European Commission that each project must have at least one international partner.

The study raises some points for a future research agenda. *First*, the estimation of this model for different FPs would shed some light on the temporal evolution of these effects and provide some insight into the integration of R&D collaborations with respect to geography and technology. A *second* promising research direction would be to apply the spatial interaction model for cross-region collaboration in different subprogrammes. This may provide insight on the role of the separation variables across different scientific fields. *Third*, the definition of the region-by-region collaboration matrix is based on the assumption that all project participations are equally important. This is of course an approximation to reality. Thus, other counting methods could be explored.

**Acknowledgements.** The work presented here is partly funded by the EU-FP6-NEST project NEMO ('Network Models, Governance and R&D collaboration networks'), contract number 028875. We thank three anonymous referees for valuable comments.

**Appendix A**

This study disaggregates Europe's territory into 255 NUTS-2 regions located in the EU-25 member states (except Cyprus and Malta) plus Norway and Switzerland. We exclude the Spanish North African territories of Ceuta y Melilla, the Portuguese non-continental territories Azores and Madeira, and the French Departments d'Outre-Mer Guadeloupe, Martinique, French Guayana and Reunion. Thus, we include the following NUTS 2 regions:

| | |
|---|---|
| *Austria*: | Burgenland; Niederösterreich; Wien; Kärnten; Steiermark; Oberösterreich; Salzburg; Tirol; Vorarlberg |
| *Belgium*: | Région de Bruxelles-Capitale/Brussels Hoofdstedelijk Gewest; Prov. Antwerpen; Prov. Limburg (BE); Prov. Oost-Vlaanderen; Prov. Vlaams-Brabant; Prov. West-Vlaanderen; Prov. Brabant Wallon; Prov. Hainaut; Prov. Liége; Prov. Luxembourg (BE); Prov. Namur |
| *Czech Republic*: | Praha, Stredni Cechy, Jihozapad, Severozapad, Severovychod, Jihovychod, Stredni Morava, Moravskoslezsko |
| *Denmark*: | Danmark |
| *Estland*: | *Eesti* |
| *Germany*: | Stuttgart; Karlsruhe; Freiburg; Tübingen; Oberbayern; Niederbayern; Oberpfalz; Oberfranken; Mittelfranken; Unterfranken; Schwaben; Berlin; Brandenburg; Bremen; Hamburg; Darmstadt; Gießen; Kassel; Mecklenburg-Vorpommern; Braunschweig; Hannover; Lüneburg; Weser-Ems; Düsseldorf; Köln; Münster; Detmold; Arnsberg; Koblenz; Trier; Rheinhessen-Pfalz; Saarland; Chemnitz; Dresden; Leipzig; Dessau; Halle; Magdeburg; Schleswig-Holstein; Thüringen |
| *Greece*: | Anatoliki Makedonia; Kentriki Makedonia; Dytiki Makedonia; Thessalia; Ipeiros; Ionia Nisia; Dytiki Ellada; Sterea Ellada; Peloponnisos; Attiki; Voreio Aigaio; Notio Aigaio; Kriti |
| *Finland*: | Itä-Suomi; Etelä-Suomi; Länsi-Suomi; Pohjois-Suomi |
| *France*: | Île de France; Champagne-Ardenne; Picardie Haute-Normandie; Centre; Basse-Normandie; Bourgogne; Nord-Pas-de-Calais; Lorraine; Alsace; Franche-Comté; Pays de la Loire; Bretagne; Poitou-Charentes; Aquitaine; Midi-Pyrénées; Limousin; Rhône-Alpes; Auvergne; Languedoc-Roussillon; Provence-Côte d'Azur; Corse |
| *Hungary*: | Kuzup-Magyarorszßg, Kuzup-Dunssnt, Nyugat-Dunssnt, Dus-Dunsst, Oszak-Magyarorszßg, Oszak-Alfald, Dus-Alfad |
| *Ireland*: | Border, Midland and Western; Southern and Eastern |
| *Italy*: | Piemonte; Valle d'Aosta; Liguria; Lombardia; Trentino-Alto Adige; Veneto; Friuli-Venezia Giulia; Emilia-Romagna; Toscana; Umbria; Marche; Lazio; Abruzzo; Molise; Campania; Puglia; Basilicata; Calabria; Sicilia; Sardegna |
| *Latvia*: | Latvia |
| *Lithuania*: | Liuteva |
| *Luxembourg*: | Luxembourg (Grand-Duché) |
| *Netherlands*: | Groningen; Friesland; Drenthe; Overijssel; Gelderland; Flevoland; Utrecht; Noord-Holland; Zuid-Holland; Zeeland; Noord-Brabant; Limburg (NL) |
| *Norway*: | Oslo og Akershus, Hedmark og Oppland, Sor-İstlandet, Agder og Rogaland, Vestlandet, Trondelag, Nord-Norge |
| *Poland*: | Lodzkie, Mazowieckie, Malopolskie, Slaskie, Lubelskie, Podkarpackie, Swietokrzyskie, Podlaskie, Wielkopolskie, Zachodniopomorskie, Lubuskie, Dolnoslaskie, Opolskie, Kujawsko-Pomorskie, Warminsko-Mazurskie, Pomorskie |
| *Portugal*: | Norte; Centro (P); Lisboa e Vale do Tejo; Alentejo |
| *Slovakia*: | Bratislavsky kraj, Zaspadny Slovensko, Stredny Slovensko, Vachodny Slovensko |
| *Slovenija*: | Slovenija |
| *Spain*: | Galicia; Asturias; Cantabria; Pais Vasco; Comunidad Foral de Navar; La Rioja; Aragón; Comunidad de Madrid; Castilla y León; Castilla-la Mancha; Extremadura; Cataluña; Comunidad Valenciana; Islas Baleares; Andalucia; Región de Murcia |
| *Sweden*: | Stockholm; Östra Mellansverige; Sydsverige; Norra Mellansverige; Mellersta Norrland; Övre Norrland; Småland med Öarna; Västsverige |
| *Switzerland:* | Region Ümanique, Espace Mittelland, Nordwestschweiz, Zürich, Ostschweiz, Zentralschweiz, Ticino |
| *United Kingdom*: | Tees Valley & Durham; Northumberland & Wear; Cumbria; Cheshire; Greater Manchester; Lancashire; Merseyside; East Riding & .Lincolnshire; North Yorkshire; South Yorkshire; West Yorkshire; Derbyshire & Nottingham; Leicestershire; Lincolnshire; Herefordshire; Shropshire & Staffordshire; West Midlands; East Anglia; Bedfordshire & Hertfordshire; Essex; Inner London; Outer London; Berkshire; Surrey; Hampshire & Isle of Wight; Kent; Gloucestershire; Dorset & Somerset; Conwall & Isles of Scilly; Devon; West Wales; East Wales; North Eastern Scotland; Eastern Scotland; South Western Scotland; Highlands and Islands; Northern Ireland |